\documentclass{aa}
\usepackage{psfig}

\newcommand{\XMM}{\mbox{\it XMM-Newton\,}}
\newcommand{\xmm}{\mbox{\it XMM\,}}
\newcommand{\Chandra}{\mbox{\it Chandra\,}}
\newcommand{\BeppoSAX}{\mbox{\it BeppoSAX\,}}

\newcommand{\ROSAT}{\mbox{\it ROSAT\,}}
\newcommand{\Einstein}{\mbox{\it Einstein\,}}

\newcommand{\Lcgs}{erg~s$^{-1}$}

\newcommand{\scinot}[2]{${#1}{\times}10^{{#2}}$}
\newcommand{\mscinot}[2]{{{#1}{\times}10^{{#2}}}}

\newcommand{\simlt}{\,{\stackrel{<}{_\sim}}\,}
\newcommand{\simgt}{\,{\stackrel{>}{_\sim}}\,}
\newcommand{\msun}{${\rm M}_\odot$}

\newcommand{\radec}[7]{$\alpha={#1}^h{#2}^m{#3}\fs{#4}$, $\delta={#5}\degr{#6}\arcmin{#7}\arcsec$}

\begin{document}
\include{jrnl_macros}

%\thesaurus{
%03 % Extragalactic astronomy
%(11.09.1 M31; % Galaxies: individual: M31
%11.19.2; % Galaxies: spiral
%11.07.1; % Galaxies: general
%11.09.4; % Galaxies: ISM
%13.25.2) % X-rays: galaxies
%13.25.3) % X-rays: general
%}

   \title{The central region of M31 observed with \XMM 
   \thanks{Based on observations obtained with XMM-Newton, an ESA
   science mission with instruments and contributions directly funded
   by ESA Member States and the USA (NASA).}}
   \subtitle{I. Group properties and diffuse emission}
   \titlerunning{M31 observed with \XMM:  I. group properties and diffuse emission}

   \author{
R.~Shirey\inst{1} \and
R.~Soria\inst{2} \and
K.~Borozdin\inst{3} \and
J.P.~Osborne\inst{4} \and
A.~Tiengo\inst{5} \and
M.~Guainazzi\inst{5} \and 
C.~Hayter\inst{4} \and
N.~La~Palombara\inst{6} \and
K.~Mason\inst{2} \and
S.~Molendi\inst{6} \and
F.~Paerels\inst{7} \and
W.~Pietsch\inst{8} \and
W.~Priedhorsky\inst{3} \and
A.M.~Read\inst{8} \and
M.G.~Watson\inst{4} \and
R.G.~West\inst{4}
}

   \authorrunning{Shirey et al.}
	
   \offprints{R.~E.~Shirey}
   \mail{shirey@xmmom.physics.ucsb.edu}

   \institute{
%1
Department of Physics, University of California, Santa Barbara, Santa Barbara, CA 93106, USA 
\and
%2
Mullard Space Science Laboratory, University College London, Holmbury St.~Mary, Dorking, RH5 6NT, UK
\and
%3
NIS-2, Space and Remote Sensing Sciences, Los Alamos National Laboratory, Los Alamos, NM 87545, USA
\and
%4
Department of Physics \& Astronomy, University of Leicester, Leicester LE1 7RH, UK
\and
%5
XMM-Newton SOC, VILSPA-ESA, Apartado 50727, 28080 Madrid, Spain
\and
%6
Istituto di Fisica Cosmica ``G.Occhialini'', Via Bassini 15, 20133, Milano, Italy
\and
%7
Columbia Astrophysics Laboratory, Columbia University, New York, NY 10027, USA
\and
%8
Max Planck Institut f\"{u}r Extraterrestrische Physik, Giessenbachstra{\ss}e, D-85741 Garching bei M\"{u}nchen, Germany
}

   \date{Received [date]; accepted [date]}

   \abstract{
We present the results of a study based on 
an \XMM\ Performance Verification observation
of the central 30\arcmin\ of the nearby spiral galaxy M31.  In the
34-ks European Photon Imaging Camera (EPIC) exposure, we detect 116
sources down to a limiting luminosity of \scinot{6}{35}~\Lcgs\
(0.3--12~keV, $d=760$~kpc).  The luminosity distribution of the sources
detected with \XMM\ flattens at luminosities below $\sim2.5 \times
10^{37}$~\Lcgs.  We make use of hardness ratios for the detected
sources in order to distinguish between classes of objects such as
super-soft sources and intrinsically hard or highly absorbed sources.
We demonstrate that the spectrum of the unresolved emission in the
bulge of M31 contains a soft excess which can be fitted with a
$\sim$0.35-keV optically-thin thermal-plasma component clearly distinct
from the composite point-source spectrum.  We suggest that this may
represent diffuse gas in the centre of M31, and we illustrate its
extent in a wavelet-deconvolved image.
}

   \maketitle
%
%________________________________________________________________

\section{Introduction}

%place this figure early to force it onto page 2 of the paper
\begin{figure*}
\vbox{\psfig{figure=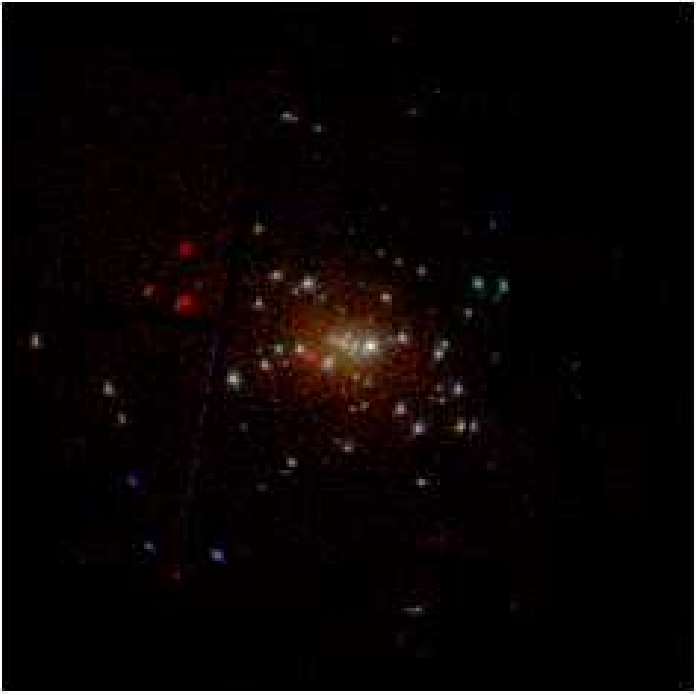,width=12cm}\vspace{-8.0cm}}
\hfill\parbox[b]{5.5cm}{
\caption{Three-colour \XMM\ EPIC MOS1 image of the central 
30\arcmin$\times$30\arcmin\ of M31.  The red, green, and blue
intensities correspond to logarithmically-scaled counts from energy
bands of 0.3--0.8~keV, 0.8--2~keV, and 2--8~keV respectively.  The
image was constructed with 2\arcsec\ pixels and has been smoothed with
a Gaussian of FWHM 4.4\arcsec, approximately equal to that of the MOS1
PSF.
\vspace{4.0cm} }
}
\label{fig:image}
\end{figure*}

Being the closest spiral galaxy to our own, the Andromeda Galaxy (M31) is
in many respects ideal for the study of X-ray emission in a galaxy
similar to the Milky Way.  The sources in M31 are observed at a nearly
uniform distance, and, owing to the inclination of the galaxy (77$^{\circ}$), 
they are viewed through a substantially lower absorption column than
for sources in the Galactic plane.
	
In a recent review, van den Bergh (2000) reports a
distance modulus to M31 of 24.4$\pm$0.1, corresponding to a distance
of 760~kpc. We adopt this value in our analysis, and for consistency
we scale to this distance when discussing published luminosities which
assume a different distance.

Over 100 X-ray sources in M31 were detected with the \Einstein\
observatory (Trinchieri \& Fabbiano 1991; van Speybroeck et al.\ 1979).  The brightest X-ray
source in M31 was found to have a luminosity of
$\sim\mscinot{3}{38}$~\Lcgs, approximately the Eddington luminosity
for spherical accretion onto a 1.4~\msun\ neutron star.  For sources
down to $\mscinot{2}{36}$~\Lcgs, the luminosity distribution was
reported to be consistent with a single power law which, extrapolated
to fainter levels, could fully account for the X-ray emission from
the bulge of M31.

Primini, Forman \& Jones (1993) detected 86 X-ray sources
in the central 34\arcmin\ of M31 with the \ROSAT\ HRI.  They found a
break in the luminosity distribution at $\sim\mscinot{2}{37}$~\Lcgs,
below which the distribution of sources flattened.  This flattening
suggested that the detected population of X-ray sources could account
for only $\sim$15--26\% of the unresolved X-ray emission in M31.
Contributions from known less-luminous populations of X-ray sources
also could not fully account for the unresolved emission, suggesting
that the remaining emission is truly diffuse or due to a new class of X-ray
sources.

In an extensive, 6.3 deg$^2$, survey of M31 with the \ROSAT\ PSPC, 396
X-ray sources were detected (Supper et al.\ 1997).  However, only 22 of
these sources were detected in the bulge region ($r < 5$\arcmin) due to the
resolution of the PSPC.

In the first \Chandra\ observation of M31, the nuclear source seen
with \Einstein\ and \ROSAT\ was resolved into five sources
(Garcia et al.\ 2000).  One of these sources is located within 1\arcsec\ of
the radio nucleus of M31 and exhibits an unusually soft X-ray
spectrum, suggesting that it may be associated with the central
super-massive black hole.  A few more pairs of previously unresolved
sources and a new transient were also detected within 30\arcsec\ of
the nucleus.

M31 was selected as an \XMM\ (Jansen et al.\ 2001) Performance Verification
target in order to demonstrate the capabilities of \xmm\ in performing
spectral and timing studies in a field of point sources and extended
emission.
In this paper we focus on the group properties of the X-ray point
sources in M31 as well as the diffuse emission.  In a companion paper
(Osborne et al.\ 20001, Paper II), we discuss the spectral and timing 
properties of individual X-ray sources in M31.

%__________________________________________________________________

\section{\XMM\ Observations}

The central region of M31 was observed with \XMM\ on 2000 July 25.
The observation was centred on the core of M31
(\radec{00}{42}{43}{0}{+41}{15}{46.0} J2000), with a field of view of
30\arcmin\ in diameter for the three European Photon Imaging Camera (EPIC)
instruments.  Exposures of 34.8~ks were obtained with
each of the two EPIC MOS instruments (Turner et al.\ 2001), and a 31.0-ks
exposure was obtained with the EPIC PN (Str\"{u}der et al.\ 2001).  All three
EPIC instruments operated in full-window mode with the medium
optical blocking filter.  The two Reflection Grating Spectrometer
(RGS) cameras (den Herder et al.\ 2001) each obtained 43-ks exposures (not
discussed here).
The Optical/UV Monitor Telescope (OM; Mason
et~al.~2001) filter wheel was set to the blocked
position during this observation; however, UV and optical exposures
with the OM are planned during upcoming Guaranteed-Time Observations
of M31.

The background rate in the EPIC detectors was steady for the entire
observation except for a background flare during the final 5~ks.  Data
from this background flare are excluded from the image in
Figure~\ref{fig:image} and from our analysis of extended regions;
however, we have found that they do not significantly affect the
spectra of individual point sources. Therefore, the entire data-set was 
used for the extraction of discrete sources.

We used the \XMM\ Science Analysis System (versions from 2000 Sept.)
to reduce the EPIC data to calibrated event lists, produce images, and
extract spectra.  We used a combination of SAS programs and external
software to further analyse the data.  In this paper, we concentrate
on data from MOS1 but we have used a sample of data from MOS2 and PN to confirm
the consistency of our results.  Slight differences in the locations of
sources on the outer MOS1 CCDs relative to MOS2 and PN indicate that
the calibration of chip boundaries must be refined. 
We used the response matrix {\tt
mos1\_medium\_all\_v3.17\_15\_tel4.rsp} for fitting MOS1
spectra; we checked that our results 
do not change significantly if the more recent response matrix 
{\tt mos1\_medium\_all\_qe17\_rmf3\_tel5\_15.rsp} (2000 Oct.) is used.

%__________________________________________________________________
\section{EPIC Images}

The EPIC images of the central 30\arcmin\ of M31 (e.g., the MOS1
image in Figure~\ref{fig:image}) contain more than 100 discrete X-ray
sources as well as unresolved emission near the centre.  
The images show four very soft (red) sources: two of them were 
identified by Kahabka (1999) as candidate
super-soft sources (SSS), based on a wider \ROSAT\ PSPC\ hardness-ratio
criteria than the original sample of SSS identified by Supper et~al.\
(1997); one is coincident with a super-nova remnant; the 
fourth one is an unidentified source detected with the \ROSAT\ HRI
(Primini et al.\ 1993).
Several very hard (blue) sources are also present.  The nature of
these extreme-colour objects is discussed below in the context of the
point-source group properties.

%__________________________________________________________________
\section{Discrete X-ray sources}

\subsection{Source Detection}
Sources were detected with the IRAF routine DAOPHOT (Stetson
1987); the SAS produced similar results.  We
determined the point spread function (PSF) from the brightest sources
in our image rather than from calibration models, and we used the
measured PSF to fit source intensities.  The FWHM of the PSF is a
function of energy and distance from the image centre; for the core
region ($r < 5$\arcmin, $\sim$1.1~kpc), where most of the sources are
found, $FWHM = $ 4\arcsec--7\arcsec.  It is more difficult to
determine the PSF for sources at $\simgt 10$\arcmin\ from the centre.
For a few of these off-axis sources, we compared the count rate found
with the PSF fitting routine with the rate obtained by considering a
30\arcsec\ source circle and a 15\arcsec\ sky annulus. We find that
the count rates obtained with the two methods are similar. We also
correct the count rates of all the sources for the vignetting of the
XMM telescope, based on the Current Calibration Files provided with
the SAS.

We detect 116 discrete sources above a $5\sigma$ threshold.
Two sources previously identified as foreground or background objects
(\ROSAT\ HRI sources 9 and 82 Primini et~al.\ 1993)
have been excluded from the analysis below.  In addition, three
sources at or near the nucleus have been resolved into two or more
sources with \Chandra\ (Garcia et al.\ 2000) and are thus also excluded from
our discussion of the luminosity distribution.

%--------------------------------------------------------------------

\subsection{Spectral Characterisation}

% Spectra of bright sources
\begin{figure}
\psfig{figure=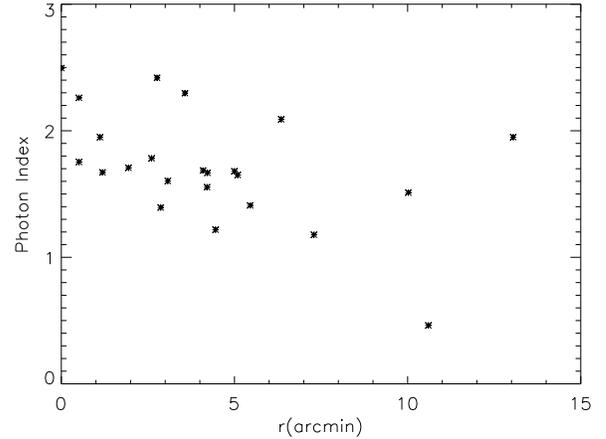,width=8.8cm,angle=90}
\caption{Photon index versus distance from the nucleus ($r$) 
for 23 bright sources with $> 500$ counts in the first 30~ks of the MOS1 
observation. (Two super-soft sources have been 
excluded from this plot, as well as two sources not belonging to M31.)}
\label{fig:photindex}
\end{figure}

We have fitted simple absorbed power-law models to the spectra of the 23 
brightest point sources (those with more than 500 counts above
the background in the first 30~ks of observation).  
We present more detailed spectral fits of selected
sources in Paper~II.  
In order to obtain the spectra of the bright sources, we used an
extraction radius of $\sim$15\arcsec; we subtracted a background
extracted from an annulus at the same distance from the galactic centre as the
source and from which the point sources have been removed.  The resulting
spectra were fitted with an absorbed power-law model, with the column
density fixed at $N_H = \mscinot{1}{21}$~cm$^{-2}$. 
Figure~\ref{fig:photindex} shows the fitted photon
indices plotted versus the distance of each source from the M31 nucleus.
We note that the hardest sources tend to
be outside the central few arcminutes of the galaxy.

%\NOTE{In the inner part of the M31 the bulge population
%dominates (LMXB) while further out one might expect more HMXB. Should
%not HMXB have harder spectra (broken power laws) then LMXB (more thermal
%brems)?} 

\begin{figure}
\psfig{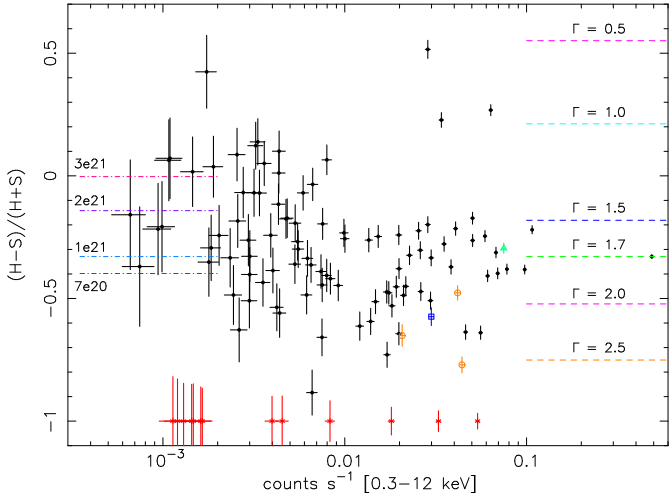}
\caption{Hardness ratio of the discrete sources detected in M31,
versus the total count rate in the $0.3$--$12$ keV band.
$S$ is the count rate in the $0.3$--$2$ keV band;
$H$ is the count rate in the $2$--$12$ keV band.
See the text for an explanation of plot symbols.  }
\label{fig:countratio}
\end{figure}

\begin{figure}
\psfig{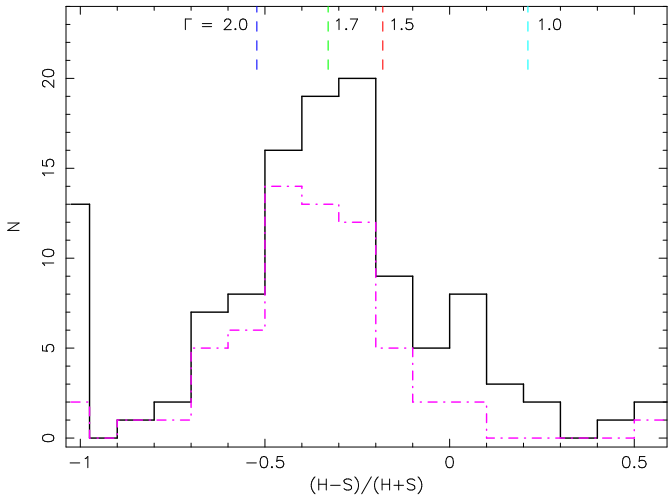}
\caption{Hardness distribution of all detected sources (solid line) and 
sources at $r < 5$\arcmin\ (dash-dotted line) compared with expected
hardness ratio for various values of the photon index $\Gamma$, at
fixed $N_H =\mscinot{1}{21}$~cm$^{-2}$.  Sources plotted with hardness
$= -1$ are not detected in the hard band.
}\label{fig:histogram}
\end{figure}

In order to characterise the gross spectral properties of all the
detected sources, including those for which we do not have enough
counts for spectral fitting, we derived source counts in soft (S) and
hard (H) energy bands of 0.3--2.0~keV and 2.0--12~keV respectively,
and we constructed hardness ratios defined as $(H-S)/(H+S)$.  In
Figure~\ref{fig:countratio}, we show the hardness ratio versus total
count rates of all the 116 sources detected in EPIC MOS1.  Thirteen
sources detected only in the soft band are plotted with a hardness
ratio $= -1$.  Among this group of very soft sources, three have an
estimated emitted luminosity $\simgt 10^{37}$ erg s$^{-1}$ (see
Section 4.3); as mentioned above, two of them are super-soft
sources, and the third one is identified with an SNR.  The three bright, blue
sources in Figure~\ref{fig:image} all have hardness ratios
significantly larger than the other sources in
Figure~\ref{fig:countratio}.  We show in Paper~II 
that these sources are intrinsically hard, rather than highly absorbed.
In Figure~\ref{fig:countratio}, we have included three core sources
resolved with \Chandra, plotted here as open circles. The softest (and
brightest) of these three unresolved
\XMM\ sources contains the true nucleus CXO J004244.2+411608
(Garcia et al.\ 2000); its hardness ratio confirms that the spectrum of the
nucleus is very soft.  The open triangle marks the bright transient
found with \Chandra\ (Garcia et al.\ 2000).  The open square marks a bright new
transient source we detected with \XMM\ (Paper~II).

On the right-hand side of Figure~\ref{fig:countratio}, we have marked
with dashed lines the expected hardness ratios of sources
characterised by a pure power-law spectrum of various photon indices
(from 0.5 to 2.5) for a fixed column density $N_H = \mscinot{1}{21}$.
On the left-hand side, we have marked with dash-dotted lines the
expected hardness ratios for various column densities $N_H =
\mscinot{7}{20}$--$\mscinot{3}{21}$ cm$^{-2}$, for a fixed photon
index $1.7$.  The majority of sources we detect have hard and soft
intensities consistent with a power law of photon index 1.0--2.5 and
column densities $N_H = \mscinot{7}{20}$--$\mscinot{3}{21}$ cm$^{-2}$.
Histograms of the hardness ratio distribution for all sources and
for those sources within 5\arcmin\ of the M31 nucleus are shown in
Figure~\ref{fig:histogram}.  The hardness ratio plots justify our
choice of a power-law spectral model with photon index 1.7 and $N_H =
\mscinot{1}{21}$ for our conversion from count rate to luminosity (see
Section~\ref{sec:lumdistr}).

\subsection{Luminosity Distribution \label{sec:lumdistr}}

\begin{figure}
\psfig{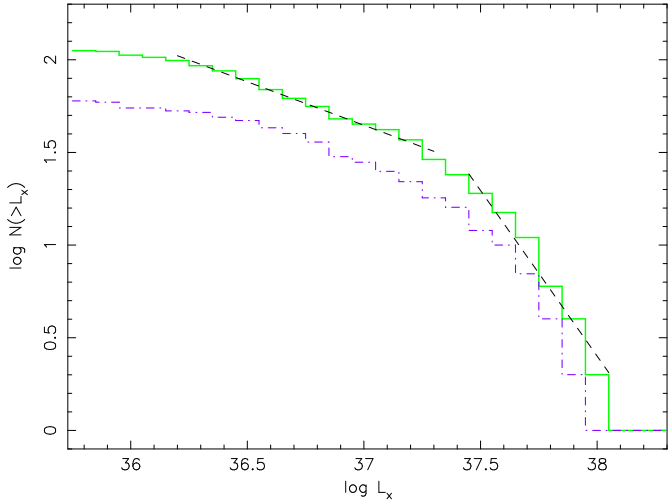}
\caption{ 
Integral luminosity distribution of X-ray sources detected with \XMM\
at 0.3--12~keV over the full 30\arcmin\ EPIC-MOS field of view (solid
line) and in the core of M31 ($r < 5$\arcmin\ from the centre;
dashed-dotted line). $L_{\rm x}$ is in units of \Lcgs.  The indices of the
power-law fits to the high- and low-luminosity parts of the total
distribution are $-1.79$ and $-0.47$ respectively (see text for
details). The unabsorbed luminosity of the $\sim 20$ brightest 
sources has been directly determined with spectral fits, 
while the values for all other sources are based on 
a count rate to luminosity conversion
factor of 1 count s$^{-1}$ $= 8.9 \times 10^{38}$ erg
s$^{-1}$. }
\label{fig:lumdistr}
\end{figure}

We convert from count rates to unabsorbed luminosities in the
$0.3$--$12$ keV range by using an absorbed power-law model with
$N_{\rm H} = \mscinot{1}{21}$ cm$^{-2}$ and $\Gamma = 1.7$, for a
distance of 760~kpc.  This gives a count rate to luminosity conversion
factor of 1 count s$^{-1}$ $= 8.9 \times 10^{38}$ erg
s$^{-1}$.
If we take an average of the 23 brightest sources (see Figure 2), 
we obtain a conversion factor of 1 count s$^{-1}$ $= (8.4 \pm 1.0) 
\times 10^{38}$ erg s$^{-1}$, consistent with the previous 
model. For the brightest source only, a conversion
factor of 1 count s$^{-1}$ $= 9.1 \times 10^{38}$ erg
s$^{-1}$ is obtained.
The observed spread in the values of the conversion factor 
for sources with different spectral characteristics gives us 
an estimate of the error ($\approx 10$\%) 
in the luminosity of faint sources 
for which no spectral fit is available.
The limiting sensitivity of our sample is $\approx 6 \times 10^{35}$ erg
s$^{-1}$.

The cumulative luminosity distribution for all the X-ray sources in
the EPIC MOS field of view is shown in Figure~\ref{fig:lumdistr}.  We
also determined a luminosity distribution for the 60 sources detected
in core of M31 ($r < 5$\arcmin\ from the centre); we plot it in
Figure~\ref{fig:lumdistr} as a dash-dotted line.  Our data show a
flattening of the luminosity distribution for $L_{\rm x}\ \simlt\ 
2.5 \times 10^{37}~$\Lcgs.  We fitted the low- and high-luminosity
sections of the total distribution with simple power laws.  For $36.2
\leq \log L_{\rm x} < 37.4$, we obtain a power-law index of $-0.47 \pm
0.03$; for $37.4 \leq \log L_{\rm x} < 38.1$, the power-law index is
$-1.79 \pm 0.26$.  For the core source distribution, the power-law
indices are $-0.43 \pm 0.04$ and $-1.77 \pm 0.35$ respectively.  Thus,
we find no significant difference in the shape of the distribution for
sources in the core compared with the distribution for all 
the sources out to $r \approx
15$\arcmin.  These results are in agreement with the luminosity
distribution found by Primini et al.\ (1993) and Supper
et~al.\ (1997).

Based on \Chandra\ deep field results (Giacconi et al.\ 2000), we estimate
that at the faint end of the luminosity distribution, at $\log L_{\rm x} =
36.2$, we might expect 10--20 background AGN in the field of view.  We
would also expect about half that number of foreground K and M stars
(following Supper et~al. 1997).  Correction for the
contribution of such foreground/background objects would further
flatten the faint end of the full-field-of-view distribution but would not
significantly affect the brighter portion of the distribution.  In the
much smaller core region ($r < 5$\arcmin), we estimate the
contribution of background AGN or foreground stars to the luminosity
distribution above $\log L_{\rm x} = 36.2$ to be minimal, i.e., in the core
we expect $\sim 1$--3 AGN and 0 or 1 stars brighter than that level.

The $0.3$--$12$ keV observed luminosity of the brightest X-ray source
in our sample is $L_{\rm x} = (3.9 \pm 0.2) \times 10^{38}$~\Lcgs,
corresponding to an emitted luminosity of $(4.5 \pm 0.2) \times
10^{38}$~\Lcgs\ for the fitted spectral model and column density. For
a comparison with earlier \ROSAT\ observations
(Primini et al.\ 1993; Supper et al.\ 1997), this corresponds to an emitted luminosity
of $(2.6 \pm 0.1) \times 10^{38}$~\Lcgs\ in the $0.1$--$2.4$ keV band,
and of $(2.9 \pm 0.1) \times 10^{38}$~\Lcgs\ in the $0.2$--$4.0$ keV
band.    These values are in agreement with the \ROSAT\
results, when we take into account our different choice of spectral
model, column density and distance to M31.

We determine an integrated $0.3$--$12$ keV emitted luminosity $L_{\rm
x} = (2.2 \pm 0.2) \times 10^{39}$~\Lcgs\ for the region at $r < 5'$.
The background was extracted from regions at $r > 7'$, excluding
detected sources.
%, and we corrected for the vignetting effect by rescaling the count rate.  
The result is insensitive to the exact model used for the spectral
fitting and the region selected for the background.  The $0.1$--$2.4$
keV emitted luminosity $L_{\rm x} = (1.35 \pm 0.15) \times
10^{39}$~\Lcgs, in agreement with the \ROSAT\ results
(Supper et al.\ 1997).  The total $0.3$--$12$ keV contribution of the
discrete sources we detected in the core, correcting for an estimated
20\% of the counts being in the PSF wings, is $L_{\rm x, s} \approx (2.0
\pm 0.1) \times 10^{39}$~\Lcgs, i.e., they account for $\sim(90 \pm 10)$\% 
of the total X-ray emission in the core.  If we extrapolate the
observed luminosity distribution of the core to a lower limit of
$10^{34}$~\Lcgs, we obtain only a small additional contribution of
$\simlt 10^{38}$~\Lcgs based on the power-law index fitted to the
distribution at $36.2 \leq \log L_{\rm x} < 37.4$, or even less if the
flattening below $\log L_{\rm x} \sim 36$ continues to fainter levels.

%--------------------------------------------------------------------

\section{Diffuse emission}

\begin{figure}
\psfig{figure=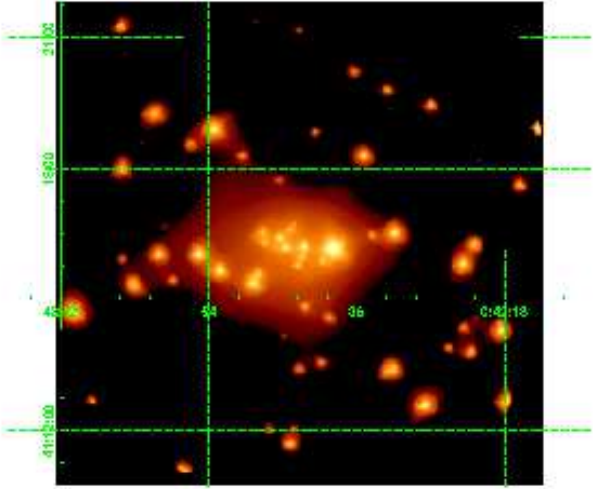,width=8.8cm,angle=0}
\caption{Wavelet-deconvolved EPIC MOS1 image of the 
central 10\arcmin\ of M31, showing diffuse emission as well as
numerous point sources.  }\label{fig:wavelet_img}
\end{figure}

\begin{figure}
\psfig{figure=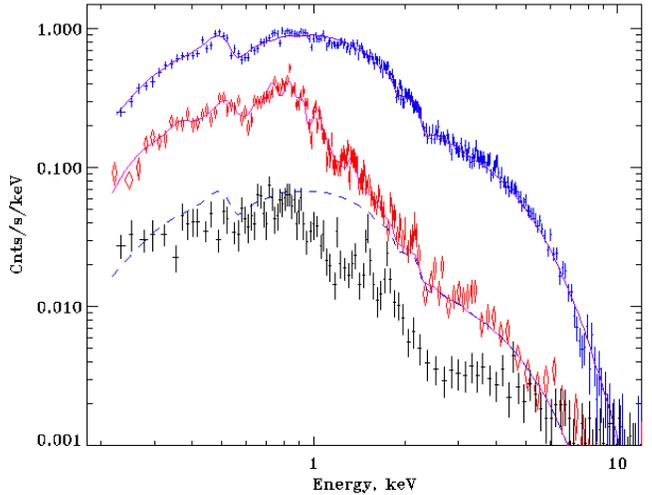,width=9.0cm,angle=270}
\caption{EPIC MOS1 spectra collected from different regions of the M31 
bulge.  Blue crosses represent the sum of 44 bright sources located
within 5\arcmin\ from the centre; the violet line is a fit of this
spectrum with a single power-law model.  The spectrum of unresolved
emission is shown by red diamonds; the blue dashed line shows
contribution of a power-law component with the same slope as the top
curve.  The need for an additional thermal plasma component associated
with the diffuse bulge emission is clearly seen. The pink solid line
is a fit of the unresolved-emission 
spectrum with a two-component model. Black crosses
below show the background, collected in the outer parts of the same MOS
chip.  All three spectra are normalised to the same area.  The two
upper spectra have been background-subtracted using the bottom
background spectrum scaled to the appropriate area.  See the text for
details of the analysis and model parameters.}
\label{fig:spec_ptsrcs_diffuse}
\end{figure}

As mentioned in the Introduction, \ROSAT\ results suggested
the presence of diffuse X-ray emission in the core of M31 
(Primini et al.\ 1993).  Recently, it was reported that \Chandra, 
in spite of its excellent imaging capabilities, does not resolve all
the soft X-ray emission from the central part of M31 into individual
sources (Garcia et al.\ 2001).  The extent of the unresolved
X-ray emission in the \XMM\ observation is illustrated by a
wavelet-deconvolved image in Figure~\ref{fig:wavelet_img}.
%However, it is difficult to calculate exactly from the image, what
%fraction of total unresolved emission can be attributed to true
%diffuse emission.  We decided to leave such analysis for future
%studies.
In order to determine the nature of the unresolved emission seen in
the image, we performed spectral analysis of several regions in the
central part of M31.
%To understand the spectrum of unresolved emission we followed an
%approach developed by Borozdin \& Priedhorsky (2000) for
%the analysis of \ROSAT/PSPC data.

%POINT-SOURCE SPECTRUM
We first extracted the composite spectrum of 44 bright point sources
detected within 5\arcmin\ of the nucleus.  Trying to maximise the
relative contribution of point sources in this spectrum, we
collected only data within small circles around each source, with
radii of 10\arcsec--30\arcsec\ ($\sim 60$--$85$\% encircled energy fraction)
depending on the source brightness (smaller radii for fainter
sources).  
%UNRESOLVED SPECTRUM
We then collected a spectrum from areas where significant
unresolved emission is present and no point sources were 
detected.  
% BACKGROUND SPECTRUM
We collected the background spectrum from the same MOS chip, but outside 
its central region, in areas where no point sources were detected.
We cannot exclude a small contamination of the background by unresolved point
sources and faint diffuse emission; however, since the background is much
fainter than the point-source and unresolved emission at all but the
highest energies (Figure 7), this is not likely to be a significant 
source of error.
The shape of the background spectrum is not significantly affected by
vignetting, because, for a given off-axis angle on the central chip, the
vignetting function varies with energy by no more than $\sim6$\% for
$E \simlt 6$~keV.  The vignetting function at different off-axis angles
on the central chip is also relatively flat, decreasing by only
$\sim$15\% from the centre to 5\arcmin\ off axis and by a total of
$\sim$25\% at 7\arcmin\ off axis.  
%

%
%The spectra are distinctively different (note background lines at
%1.485~keV and 1.74~keV, and a hard X-ray background component, none of
%which is present in the background-subtracted spectra).  
%

We scaled the spectra of the unresolved emission and of the background to
the area of the point-source spectrum.  We show the
background-subtracted point-source and unresolved spectra, along with
the background itself, in Figure~\ref{fig:spec_ptsrcs_diffuse}.
%
% POINT-SOURCE FIT
The point-source spectrum is well fitted with a power law of photon index
$1.82 \pm 0.03$ and absorption column density $N_H = (6.7 \pm 0.4)
\times 10^{20}$ cm$^{-2}$.  This spectrum is typical of low-mass X-ray 
binaries (LMXBs) and the
column density is close to the Galactic column along the line of sight
to M31.
% UNRESOLVED FIT
In contrast, the spectrum of the unresolved emission cannot be fitted 
with a single power-law component. Above $\sim$\,2~keV, a power-law
component of the same slope as the point-source spectrum is dominant,
but a significant soft excess is clearly present below this energy.
This soft excess can be fitted with an additional component from an
optically-thin thermal plasma (MEKAL) with $kT\sim0.35$~keV.

Based on the portion of the PSF that fell outside our exclusion region
around each point source, we conclude that the 
power-law component in the spectrum of the unresolved emission is
mostly due to point-source counts in the wings of the PSF. 
A further contribution to this component may come from 
faint LMXBs below our detection limit. 
An accurate determination of the relative contribution of PSF wings,
unresolved LMXBs, background, and any other component to the hard
portion of the unresolved emission will require further knowledge of
the instrument performance.  We thus defer this task to a future work.

We now turn our attention to the soft portion of the unresolved emission. 
The fraction
of encircled energy inside a chosen angular radius falls with energy for
\xmm, so that the spectrum of the PSF wings should be harder than the
extracted point-source spectrum; thus, the PSF wings cannot
contribute to the soft excess in the unresolved emission.  
Moreover, Figure 7 shows that the contribution of the background 
in this spectral region is much smaller than the soft excess.
Therefore, the soft excess and its
thermal-plasma spectrum cannot be explained by instrumental or
background effects.

What is the physical nature of this component?  
It was suggested by Irwin \& Bregman (1999)
that LMXBs could contribute to the soft excess.
However, we have not detected a significant soft component 
in the LMXB-dominated point-source spectrum. 
Furthermore,
we have shown in Fig.~\ref{fig:countratio} that the fainter sources we
detect have a similar or higher hardness ratio than most of the 
brighter sources.
Therefore, we see no evidence that LMXBs might
significantly contribute to the soft spectral excess.

We cannot exclude that a population of faint point sources 
with soft X-ray spectra, different
from the population of our detected bright sources, may be
responsible for part or all of the soft X-ray excess.  However, because the
soft excess in the unresolved emission spectrum is well fitted with a
thin thermal plasma model with significant line emission, we favour the
interpretation of the soft excess 
as truly diffuse emission from hot, optically-thin plasma.  

We fit the background-subtracted spectrum of the total emission within
$r<5$\arcmin\ using the thermal-plasma plus power-law model and found
that the thermal component contributes $\sim$10\% of the total unabsorbed
spectral flux (0.3--12 keV).  We accept this spectrally-determined
value as our best estimate of the diffuse-emission contribution to the
X-ray luminosity of the bulge of M31. This value is also consistent 
with our estimate based the discrete source distribution (Section 4.3).  
We leave a more precise
determination of the relative contribution of diffuse emission to the
total X-ray flux to a future work when calibration details are more
secure.
%__________________________________________________________________

\section{Conclusions}

Using \XMM\ data, we have confirmed that 
the X-ray emission from the bulge of M31
is dominated by bright point-like sources, most of which are likely
to be low-mass X-ray binaries.  For sources in the central region of M31, we
have confirmed that the luminosity distribution is flatter toward
lower luminosities (Primini et al.\ 1993).  The steepening of the luminosity
distribution above $2.5 \times 10^{37}$\Lcgs\ is indicative of a lack of
bright sources in M31 (cf.\ the source distribution in M33, Long et al.\ 1996).
Only two sources in our sample have a $0.3$--$12$ keV unabsorbed luminosity
$\geq 10^{38}$\Lcgs.

As in previous observations with \Einstein\ and \ROSAT\
(Trinchieri \& Fabbiano 1991; Primini et al.\ 1993), significant unresolved emission was
found to contribute to the total emission of the bulge.
The flattening of the luminosity distribution
for fainter sources means that an extrapolation of the detected 
population of point sources at lower energies 
cannot account for the total core emission
of M31.

A soft excess in the spectrum of the M31 bulge was previously
reported in \ROSAT\ and \BeppoSAX\ observations 
(Irwin \& Bregman 1999; Trinchieri et al.\ 1999).  Our analysis 
of the \XMM\ data shows that the soft component in the spectrum of the
bulge is associated with unresolved emission; this confirms
the results of Borozdin \& Priedhorsky (2000) based on
\ROSAT\ data.  

More importantly, our \XMM\
study has revealed for the first time that while the integrated
spectrum of point-like sources is featureless, the spectrum of the 
unresolved emission shows multiple emission lines typically found in the
spectrum of hot, optically thin plasma.  
Therefore, we suggest that the second
significant source of X-ray emission in the bulge is truly diffuse gas
with an effective temperature $\sim$0.35 keV.  The contribution of this
gas to the total unabsorbed X-ray luminosity is estimated to be $\sim$10\%
in 0.3--12~keV band (corresponding to $\approx 2 \times 10^{38}$\Lcgs), 
but more than 20\% in the \ROSAT\ band
(0.1--2.4~keV).  The significance of this result goes far beyond the
case of M31, because the bulge of this galaxy is often considered as a
prototype for the population of early-type X-ray galaxies.  
For example, Sarazin et al.~(2000) recently reported
that, according to \Chandra\ observations, 
23\% of X-ray emission from NGC\,4697 is
emitted by interstellar gas, contrary to their previous expectations
(Irwin et al.\ 2000).

Two more \XMM\ observations of the central region of M31 are scheduled
as part of the Guaranteed-Time program, as are observations of five
additional fields along the disk of M31.  These will allow us to reach
fainter flux levels in the bulge and to study the
populations of X-ray sources in different parts of the M31 galaxy.

%__________________________________________________________________
\begin{acknowledgements}
We thank all the members of the \XMM\ teams for their work building,
operating, and calibrating the powerful suite of instruments on-board.
\end{acknowledgements}

%\bibliography{aamnem99,m31,xmm}
%\bibliographystyle{aabib99}

\end{document}